\documentclass[preprint,12pt]{elsarticle}

\usepackage{amssymb}
\usepackage{color}
\journal{Chaos, Solitons and Fractals for publication}

\begin{document}

\begin{frontmatter}



\title{Coupled dynamics of mobility and pattern formation in optional public goods games}


\author{Li-Xin Zhong$^a$}\ead{zlxxwj@163.com}
\author {Wen-Juan Xu$^a$}
\author {Yong-Dong Shi$^{a,b}$}
\author {Tian Qiu$^c$}

\address[label1]{School of Finance, Zhejiang University of Finance and economics, Hangzhou, 310018, China}
\address[label2]{School of Finance and Research Center of Applied Finance, Dongbei University of Finance and Economics, Dalian, 116025, China}
\address[label3]{School of Information Engineering, Nanchang Hangkong University, Nanchang, 330063, China}

\begin{abstract}
In a static environment, optional participation and a local agglomeration of cooperators are found to be beneficial for the occurrence and maintenance of cooperation. In the optional public goods game, the rock-scissors-paper cycles of different strategies yield oscillatory cooperation but not stable cooperation. In this paper, by incorporating population density and individual mobility into the spatial optional public goods game, we study the coevolutionary dynamics of strategy updating and benefit-seeking migration. With low population density and slow movement, an optimal level of cooperation is easy to be reached. An increase in population density and speed-up of free-floating of competitive agents will suppress cooperation. A log-log relation between the levels of cooperation and the free-floating probability is found. Theoretical analysis indicates that the decrease of cooperator frequency in the present model should result from the increased interactions between different agents, which may originate from the increased cluster size or the speed-up of random-movement.
\end{abstract}


\end{frontmatter}

% \linenumbers

\section{Introduction}
\label{sec:introduction}
Social dilemmas describe conflict situations existing between a rational individual maximizing its own benefit and a social group pursuing collective wellbeing\cite{segbroeck,graser}. For example, as a sheepherder enjoys herding in the public greenland, the joint effort to achieve environmental sustainability suffers a heavy blow. The gas emission from a factory promoting its prosperity makes the greenhouse problem become more serious, which in turn does harm to the further development of the factory. As everyone faces the temptation to exploit the public goods and make no contribution to society, the immense benefit that can only be got through mutual cooperation becomes unattainable. This poses a challenging problem about how the individual selfishness can lead to the occurrence and maintenance of cooperation commonly found in reality. To answer this question, various mechanisms have been introduced by scientists since Darwin\cite{sigmund,perc1,wang1,szolnoki}.

Most studies on the evolution of cooperation among selfish individuals are based on game models, the spatial ultimatum game\cite{szolnoki1}, the Prisoner's Dilemma game(PDG)\cite{wang2,yun,jiang,chen}, the snowdrift game (SG)\cite{zhong1,li,wang3,zhong2,zheng}, and the public goods game (PGG)\cite{santos,perc2,zhang}, to name just a few. The iterated PDG models the interactions between two agents, in which one's contribution favors the other but not itself. Although the total income would be the highest if they both cooperate, each agent tends to defect to maximize its own profit. Therefore, the Nash equilibrium is to defect in all rounds. The PGG is an extension of the PDG to an arbitrary number of agents. In the original PGG, a group of individuals have the choice whether to make an investment into a common pool or not. An equal division of returns irrespective of one's contribution results in the situation where the defectors have an advantage over the cooperators and defection becomes a dominant strategy. To refrain from getting stuck in the deadlocks of mutual defection, a third strategy, termed a loner's strategy\cite{hauert1,rand,xu1,sasaki}, has been introduced into the original public goods game.

In the public goods game with loners, also known as the optional public goods game (OPGG)\cite{wang4}, the agents have an option whether to participate in the public goods game or not. Those who join the public goods game get a cooperator's payoff $P_{C}$ or a defector's payoff $P_{D}$, and those who do not join the public goods game get a loner's payoff $P_{L}$. Because a loner only gets a small but fixed payoff, it can win over a group of defectors but will be defeated by a group of cooperators. Therefore, an endless rock-scissors-paper cycle occurs: Loners will invade a population of defectors with fewer cooperators. Cooperators will thrive in a population of loners. A population of cooperators will be intruded by defectors.

Over the past decade, advances in statistical physics have fueled great interests in constructing a theory of complexity \cite{grigolini1}, among which researches on the three-state systems have attained great achievements. The voluntary prisoner's dilemma, the rock-scissors-paper game, the cyclic predator-prey model, the three-state Potts model and the three-state cyclic voter model are commonly used models\cite{szabo1,szolnoki2,perc3,szabo2,szabo3,szolnoki3,szabo4}. Depending upon pair approximations and mean-field theories, Szabo et al. have theoretically analyzed the cyclic dominance in evolutionary dynamics. It has been found that the symmetric solution of mean-field approximation is stable, but it is not asymptotically stable\cite{szabo5}.  As to the stationary solutions of the pair approximation, it has been found that they are unstable for small perturbations, which can be eliminated by using the four- and nine-site approximations\cite{szabo2}.

In the optional public goods game, although the existence of loners keeps the cooperators from being doomed, the cyclic oscillation of the three strategies indicates that a higher and stable level of cooperation is difficult to be reached in such a system. But in real world, cooperation is often the dominant strategy in animal and human activities and an environmental change will lead to the occurrence of different levels of cooperation. The environmental conditions include the structured space\cite{johnson,zhong3,lu,xu2,szabo,amritkara}, the population density\cite{cressman,rankin,wang5,wakano}, and the mobility of the individuals\cite{gonzalez,peruani,song,cardillo}. To find out the mechanisms determining the occurrence of different levels of cooperation in real world, it requires generalizing the model presented by Hauert et al. and incorporating the environmental conditions into the original OPGG \cite{hauert1}.

In natural and human society,  the linkage between the agents may dynamically evolve\cite{turalska,vanni}, the structured space is not fully occupied and random and purposive movements often occur\cite{helbing,meloni,droz,perc4}. Such as the migration of birds, the floating of a boat and the motion of a train. Similar dynamic processes have been found in diffusion systems\cite{grigolini2,scafetta}. X.Chen et al. have studied the role of risk-driven migration in the evolution of cooperation\cite{chen1}. Z.Wang et al. have investigated the impact of population density on the evolution of cooperation based on different game models\cite{wang5,wang6}. Z.H.Liu has studied the influence of population density and individual mobility on epidemic spreading\cite{liu}. C.P.Roca et al. and J.Y.Wakano et al. have studied the roles of mobility in the improvement of cooperation in the PGG and in the ecological PGG respectively\cite{roca,wakano}. Related studies have shown that, in the structured space, the percolation threshold plays a quite important role in the widespread of cooperation and the outbreak of diseases\cite{chen1,wang5,wang6,liu}. Motivated by the work done in partially occupied lattices, in this paper, we incorporate population density and individual mobility into the original OPGG introduced in ref.\cite{hauert1} and play the game in a square lattice with Moore neighborhood. The main findings of the study are as follows:

(1) With a predefined free-floating probability $\xi_0$, the cooperator frequency $f_C$ is determined by population density $\rho$. There exists a transition point $\rho_{tr}$, below which $f_C$ increases with the rise of $\rho$ while above which $f_C$ decreases with the rise of $\rho$. With a predefined $\rho$, $f_C$ is determined by $\xi_0$. Increasing $\xi_0$ leads to a monotonic decrease of $f_C$.

(2) Considering the size distribution $P(S)$ of individual components, we find that the power-law relation between $f_C$ and $\xi_0$ is related to the occurrence of a giant component $S_{max}\sim N$. Before the giant component occurs, as $\rho$ increases, $P(S)$ changes from an exponential to a power-law distribution and the slop of $f_C$ as a function of $\xi_0$ decreases with the rise of $\rho$. After the giant component occurs, as $\rho$ increases, $P(S)$ changes little and the slope of $f_C$ as a function of $\xi_0$ also changes little with the rise of $\rho$.

(3)As we keep an eye on the decrease of cooperation in the present model, the effect of increasing $\xi_0$ with a fixed $\rho$ is similar to the effect of increasing $\rho$ with a fixed $\xi_0$. A theoretical analysis shows that the change of the frequencies of different strategies in the present model should be determined by possible collisions between the agents with different strategies. Both the increase in $\rho$ and the increase in $\xi_0$ would result in more collisions. The more the collisions between the agents, the lower the levels of cooperation.

The paper will proceed as follows. In Section 2, we introduce the optional public goods game with purposive and random movements in a spatial setting. In Section 3, simulation results about the evolution of competitive strategies and the local agglomeration of individuals are presented and the relationship between them is discussed. In Section 4, the extinction thresholds are analyzed with mean field theory, and the possible relations between the levels of cooperation and the individual collisions are described theoretically. Section 5 summarizes the paper and gives an outlook for future studies.

\section{The model}
\label{sec:model}
A population of $N$ agents is distributed over a square lattice with side length $X$ and the Moore neighborhood (i.e., degree $k =8$), each agent on each site. For $N\leq X^2$, the population density is defined as $\rho=\frac{N}{X^2}$. For $\rho<1$, an agent can move around and occupy the firstly found empty site. In the present model, there exist two coevolutionary processes: the change of personal strategies and the movement of the agents. Once the initial position and the adopted strategy for each agent are set, the system will evolve as follows.

In the evolution of personal strategies. Initially, there exist three kinds of agents: cooperators (C), defectors (D) and loners (L). At each Monte Carlo step (MCS), firstly, each agent interacts with its nearest neighbors and gets a payoff $P_C$ (for cooperators), $P_D$ (for defectors) or $P_L$ (for loners). Owing to the neighboring restriction, the interaction group in the present model should be $n\leq 9$. Assuming $n=n_C+n_D+n_L$, in which $n_C$, $n_D$, $n_L$ represent the numbers of cooperators, defectors and loners in the interaction group respectively, the payoffs for the agents with different strategies are

\begin{equation}
P_C=\frac{rn_C}{n_C+n_D}-1,
\end{equation}

\begin{equation}
P_D=\frac{rn_C}{n_C+n_D},
\end{equation}

\begin{equation}
P_L=\sigma,
\end{equation}
in which $r$ ($>1$) is the multiplication rate and $\sigma$ generally satisfies $0<\sigma<r-1$. Therefore, a loner's payoff should be lower than the payoff of the agents in a group of cooperators and higher than the payoff of the agents in a group of defectors. If $n_C+n_D=1$, the cooperator or defector will get a loner's payoff. After all the agents have attained their payoffs,  they will make decisions on whether they should update their strategies or not. In the updating process, an agent i compares its payoff with a randomly chosen neighbor j's and adopts j's strategy with probability

\begin{equation}
\Gamma_{i\gets j}=\frac{1}{1+e^{{(P_i-P_j+\tau)}/{\kappa}}},
\end{equation}
in which $\tau$ and $\kappa$ represent the cost of strategy change and the environmental noise respectively.

The above updating mechanism shows that, in an interaction group, the evolution of the mixed strategy profile is determined by the scores of different strategies. Even if the group size is constant, the number of competitive agents, i.e. cooperators and defectors, will vary with time. Such that the cyclic oscillations will occur in a fully-occupied network setting.

In relation to the mobility of individuals. For a loner, because it only relies on a small but fixed payoff and gets nothing from a benefit-seeking competition, it is not so attractive to stay at its current site and at each time step it will move randomly with probability $\xi_L$. For cooperators and defectors, the probabilities of leaving the current sites are determined by whether they are satisfied with the present situation and the possible free-floating, which are described by the purposive-moving probability $\xi_g$ and the free-floating probability $\xi_0$ respectively. The value of $\xi_0$ is predefined and the value of $\xi_g$ is determined by the wealth of an agent, which is a cumulative payoff in the latest T MCS. For an agent i, its wealth is

\begin{equation}
w_i(t)=\sum_{t'=t-T+1}^{t} p_i(t').
\end{equation}
For a cooperator or a defector, if its wealth $w$ is less than 0, with probability

\begin{equation}
\xi_{CD}=\xi_{g}+\xi_{0},
\end{equation}
in which $\xi_{g}=1-e^{w}$, it leaves its current site and finds an empty site to stay on. In the present model, to refrain from the random disappearance of loners, at every MCS, the competitive agents will occasionally become loners with a small random-flipping probability $\varepsilon_L$.

A Monte Carlo Step can be summarized as follows. (1) Each agent interacts with its nearest neighbor(s) and gets its wealth $w$. (2) For a cooperator or a defector, if its wealth is less than 0, it leaves its current site with probability $\xi_{CD}$. If its wealth is greater than or equal to 0, it leaves its current site with probability $\xi_{0}$. For a loner, it leaves its current site with probability $\xi_L$ no matter how rich it is. The moving agents walk randomly and stay on the firstly found empty sites. For all the agents, they move sequentially. (3) The wealth of a moving agent is reset to 0 and it will not move in the next T-1 MCS. (4) Each agent interacts with its nearest neighbor(s) and gets its payoff $P$. (5) Each agent i compares its payoff with a randomly chosen neighbor j's and adopts neighbor j's strategy with probability $\Gamma_{i\gets j}$. For all the agents, they update their strategies synchronously. (6) Cooperators and defectors become loners with probability $\varepsilon_L$.

Therefore, in the present model, there are two mechanisms which determine the evolution of cooperation. The payoff determines the replicator dynamics while the wealth determines the mobility dynamics. In relation to the replicator dynamics. Different from many other PGG models where the agents collect income simultaneously from several public goods games\cite{szabo5,perc5,szolnoki4}, in the present model, the agents collect income from a single public goods game. Our concentration is how the evolutionary dynamics in a three-strategy game is affected by the population density and the free-floating probability. In relation to the local agglomeration. In the present model, increasing population density has a great influence on the size distribution of individual components, which is similar to that in the percolation problem. In the physical world, percolation theory is commonly used to explain connectivity and transport problems. For example, occupy the sites on a square lattice with probability p. For a small $p<p_c$, only small isolated clusters, which means a set of neighboring sites occupied, are observed. Increasing p leads to the growth and merging of clusters. For $p=p_c$, one dominant cluster (infinitely large cluster) occurs and at p=1 all the sites are occupied. The critical point $p_c$, at which a dominant cluster suddenly occurs, is known as the percolation threshold. The exact value of the threshold and the system property close to $p_c$ are both fundamental problems in percolation theory, which have been widely studied by physicists\cite{szabo5,stanley,gyure}. In the present model, the cluster of sites occupied by agglomerated individuals, which is called individual component throughout the paper, should be affected by population density. Both the size of the largest component and the distribution of the component sizes can effectively reflect the percolation properties. The relationship between the levels of cooperation and the occurrence of a dominant component is a favorite of ours.

\section{Results and discussions}
\label{sec:results}

In this section, we will focus on the roles of population density and free-floating probability $\xi_{0}$ in the change of cooperation. Following the work done in ref.\cite{hauert1}, in Monte Carlo simulations we choose the loner's payoff $\sigma=1$, the cost of strategy change $\tau=0.1$ and the environmental noise $\kappa=0.1$. Throughout the paper, the following parameters are also predefined: the size of the square lattice $M=X\times X=200\times 200$, the random-moving probability of loners $\xi_L=0.1$ and the random-flipping $C\to L$ (or $D\to L$) probability $\varepsilon_L=0.001$.

\begin{figure}
\includegraphics[width=12cm]{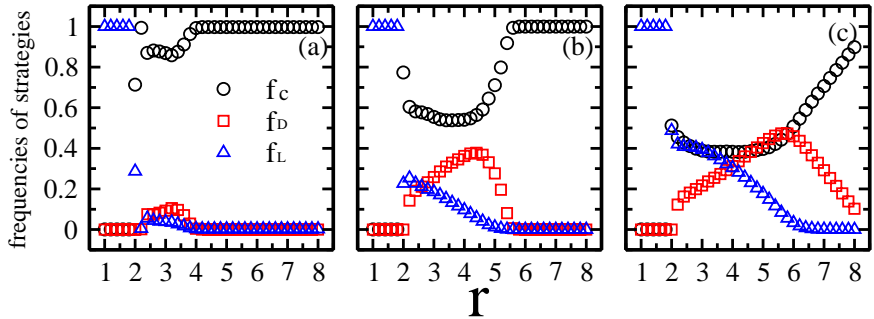}% Here is how to import EPS art
\caption{\label{fig:epsart}Frequencies of cooperators (circles), defectors (squares), and loners (triangles) as a function of multiplication rate $r$ in a square lattice with size $M=X\times X=200\times 200$ and Moore neighborhood. Other parameters are $\sigma=1$, $\tau=\kappa=0.1$, $\xi_L=0.1$, $\xi_0=0.01$, $\varepsilon_L=0.001$ and (a)$\rho=0.3$, (b)$\rho=0.5$, (c)$\rho=0.8$. The results are obtained by averaging over 10 runs and 1000 MCS after 10000 relaxation MCS in each run.}
\end{figure}

Figure 1 shows the frequencies of cooperators, defectors and loners as a function of $r$ in a square lattice with different population densities $\rho=0.3$, 0.5 and 0.8. As what has been found in a fully-occupied regular network\cite{szabo}, there exist two extinction thresholds $r_D$ and $r_L$. For $r<r_D$, loners perform better than cooperators and defectors so that all the agents become loners in the final steady state. For $r_D<r<r_L$, the three strategies coexist. The rise of $\rho$ is found to be beneficial for defectors and loners, but not for cooperators. For $r>r_L$, the survival of cooperators is favored and the loners go into extinction.

In all the three cases with different $\rho$, the values of the extinction threshold $r_D$ are the same, $r_D=\sigma+1=2$, which is also the same as that in a fully-occupied regular network\cite{hauert2}. Such a result comes from the fact that, with a small multiplication rate $r<2$, even if all the competitive agents are cooperators, the payoff of a cooperator, $P_C=r-1$, is less than a loner's payoff $\sigma=1$. Therefore, the rock-scissors-paper cycle does not occur and the system is stuck in the state where all the agents are loners. However, as we consider the extinction threshold $r_L$, we find it increases with the rise of $\rho$. For $\rho=0.3$, $r_L\sim 4$. For $\rho=0.5$, $r_L\sim 5.5$. For $\rho=0.8$, $r_L\sim 6.5$.

\begin{figure}
\includegraphics[width=12cm]{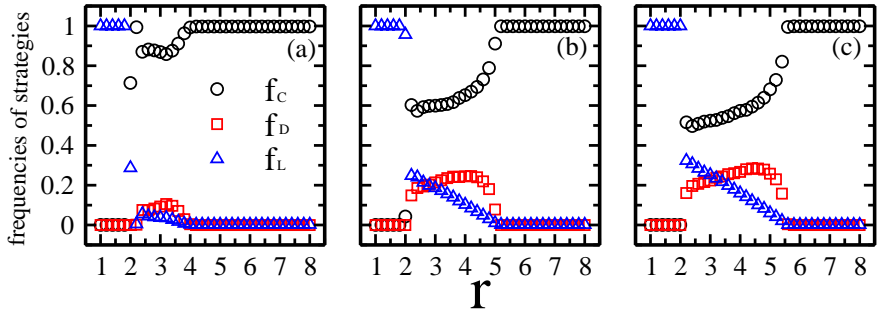}% Here is how to import EPS art
\caption{\label{fig:epsart}Frequencies of cooperators (circles), defectors (squares), and loners (triangles) as a function of multiplication rate $r$ in a square lattice with $\rho=0.3$ and (a)$\xi_0=0.01$, (b)$\xi_0=0.05$, (c)$\xi_0=0.1$. All the other parameters are the same as those in fig.1.}
\end{figure}

The rise of population density can effectively facilitate the interactions between two agents, which is more possible in the system where the agents can move freely. In fig.2, we give the frequencies of different strategies as a function of $r$ with predefined population density $\rho=0.3$ and different $\xi_0$. Comparing the results in fig.2 with those in fig.1, we find that, as to the change of the frequencies of different strategies, the effect of increasing $\xi_0$ for a fixed $\rho$ is similar to the effect of increasing $\rho$ for a fixed $\xi_0(\neq 0)$. As $\xi_0$ increases, the cooperator frequency $f_C$ decreases while the defector frequency $f_D$ and the loner frequency $f_L$ increase within a large range of $r$. As $\xi_0$ increases from 0.01 through 0.05 to 0.1, the extinction threshold $r_L$ increases from 4 through 5 to 5.5 accordingly.

The results in fig.1 and fig.2 indicate that the increase in the meeting probability will lead to the decrease of cooperation, regardless of its coming from the rise of population density or the rise of free-floating probability.

\begin{figure}
\includegraphics[width=6cm]{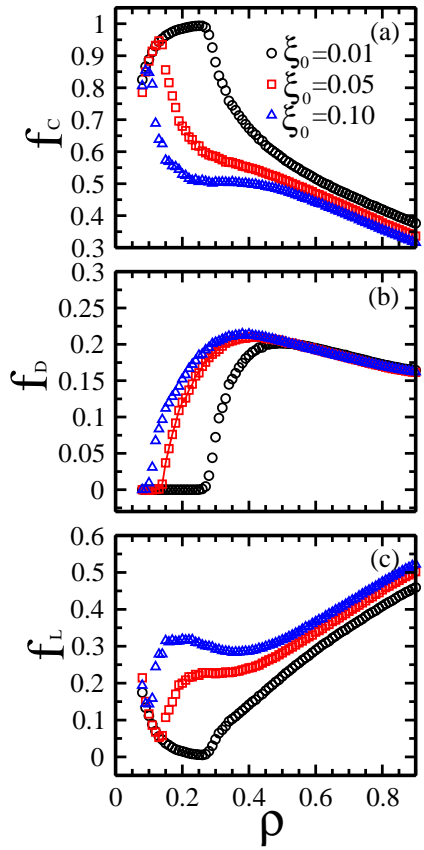}% Here is how to import EPS art
\caption{\label{fig:epsart}Frequencies of (a) cooperators, (b) defectors, and (c) loners as a function of population density $\rho$ for $r=2.5$ and different $\xi_0=0.01$(circles), 0.05(squares), 0.1(triangles). All the other parameters are the same as those in fig.1.}
\end{figure}

It is instructive to ask:  can we draw a conclusion that the fewer the chances to meet, the higher the levels of cooperation in the present model? To examine whether an optimal level of population density exists, in fig. 3 we give the frequencies of the three strategies as a function of $\rho$ for different $\xi_0$. With a small and fixed $\xi_0=0.01$, there exists a transition point $\rho_{tr}$. For $\rho<\rho_{tr}$, defectors do not exist in the population. The number of cooperators increases while the number of loners decreases with the rise of $\rho$. For $\rho>\rho_{tr}$, the number of cooperators decreases while the number of loners increases with the rise of $\rho$. The number of defectors firstly increases and then decreases with the rise of $\rho$. Increasing the free-floating probability leads to the decrease of the levels of cooperation but not the disappearance of the transition point. As $\xi_0$ ranges from 0.01 to 0.1, the transition point $\rho_{tr}$ changes from 0.28 to 0.1 and the cooperator frequency at the transition point decreases from 1 to 0.86 accordingly. The existence of the transition point indicates that, there exists an optimal level of population density, with which the system will reach the highest level of cooperation.

\begin{figure}
\includegraphics[width=12cm]{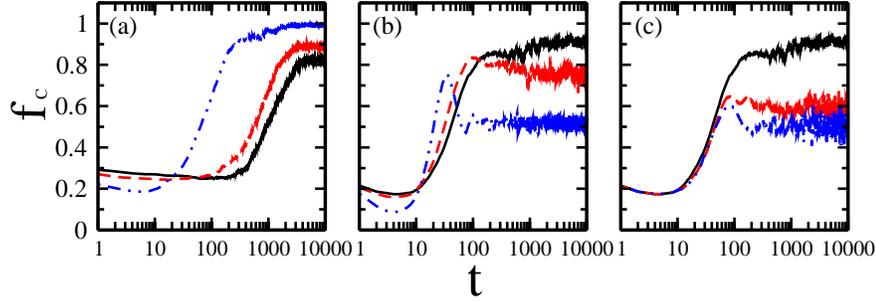}% Here is how to import EPS art
\caption{\label{fig:epsart}Time-dependent frequencies of cooperators with r=2.5, $\xi_L=0.1$, $\varepsilon_L=0.001$ and (a) $\xi_0=0.01$, $\rho=0.07$(black), 0.10(red), 0.25(blue); (b)$\xi_0=0.01$, $\rho=0.30$(black), 0.35(red), 0.60(blue); (c) $\rho=0.3$, $\xi_0=0.01$(black), 0.05(red), 0.10(blue).}
\end{figure}

To have a close eye on the evolution of cooperation below and above the transition point $\rho_{tr}$, in fig. 4(a) and (b) we plot the time-dependent frequencies of cooperators for different $\rho$. For comparison, we also plot the time-dependent cooperator frequencies for different $\xi_0$ in fig. 4(c). For $\rho<\rho_{tr}$, as the time passes $f_C$ becomes stable. The change of $\rho$ only leads to the change of the average value of cooperator frequencies but not the fluctuations of $f_C$. For $\rho>\rho_{tr}$, the change of $\rho$ not only results in the change of $f_C$  but also the stability of cooperation. Increasing $\rho$ leads to large fluctuations of the levels of cooperation. Figure 4(c) shows increasing $\xi_0$ also leads to large fluctuations of the levels of cooperation. Such results indicate that, whether the decrease of cooperator frequency results from the increase in $\rho$ or the increase in $\xi_0$, the large fluctuations of the strategies are detrimental to cooperation.

\begin{figure}
\includegraphics[width=5cm]{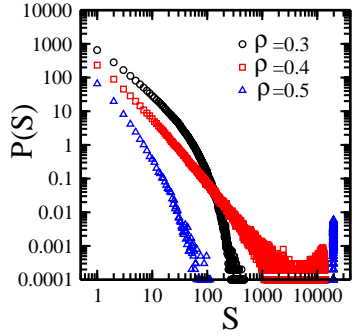}% Here is how to import EPS art
\caption{\label{fig:epsart}The size distribution of individual components in the evolved system with r=2.5, $\xi_L=0.1$, $\varepsilon_L=0.001$ and different population density $\rho=0.3$(circles), 0.4(squares), 0.5(triangles).}
\end{figure}

To find the relationship between the frequencies of strategies and the population patterns, in fig.5 we plot the size distribution of individual components for different $\rho$. As $\rho$ increases from 0.3 through 0.4 to 0.5, the size distribution changes from an exponential distribution through a power-law distribution to a giant component accompanied by a power-law distribution.

\begin{figure}
\includegraphics[width=5cm]{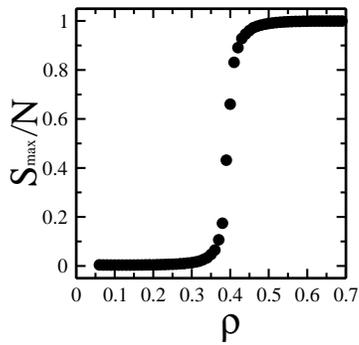}% Here is how to import EPS art
\caption{\label{fig:epsart}The size of the largest component as a function of population density $\rho$ in the evolved system with r=2.5, $\xi_L=0.1$ and $\varepsilon_L=0.001$.}
\end{figure}

In fig.6 we display the size of the largest component as a function of $\rho$. As $\rho$ ranges from 0.1 to 0.3, the size of the largest component has little change with the rise of $\rho$. As $\rho$ ranges from 0.3 to 0.5, the size of the largest component has a sharp increase with the rise of $\rho$. For $\rho>0.5$, nearly all the agents are in the same component and the largest component changes little with the rise of $\rho$.

Comparing the results in fig.3 with the results in fig.5 and fig.6 we find that, for low population density, the agents are scattered and the levels of cooperation are somewhat high. For high population density, nearly all the agents are in the same component and the levels of cooperation become lower. Therefore, the results in fig.3 can be understood as follows: For quite low population density, the dissatisfied agents' leave makes it impossible for the defectors to exploit the cooperators and the group of defectors will finally be doomed by the random-moving loners. And for the cooperators who leave the competing group, because of the low population density, it is not easy for them to find another cooperator to collaborate with, and they will finally become solitary agents whose payoffs are equal to the payoff of a loner. The occasional $C\to L$ (or $D\to L$) random-flipping can protect the loners from extinction. Therefore, under such an environment, only cooperators and loners are kept in the final steady state. But for high population density or a high free-floating probability, it is easy for defectors to intrude into cooperator clusters and the average level of cooperation will accordingly decrease with the rise of $\rho$ or the rise of $\xi_0$.

\begin{figure}
\includegraphics[width=5cm]{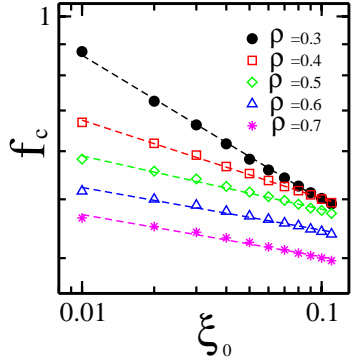}% Here is how to import EPS art
\caption{\label{fig:epsart}Log-log plot of the cooperator frequency $f_C$ vs $\xi_0$ with $\rho=0.3$, 0.4, 0.5, 0.6, 0.7. All the other parameters are r=2.5, $\xi_L=0.1$ and $\varepsilon_L=0.001$ . The fitted lines satisfy the equation $f_C=a\xi_0^{-b}$.}
\end{figure}

To have a deep understanding of the roles of free-floating in the increase or decrease of cooperation, in fig.7 we plot $f_C$ vs $\xi_0$ for different $\rho$. From fig.7 we find that, for low population density, the frequency of cooperators decreases sharply with the rise of $\xi_0$. The rise of $\rho$ makes such a changing tendency become ease. As we fit curves to data points in fig.7, it is found that $f_C$  and $\xi_0$ satisfy the equation $f_C=a\xi_0^{-b}$, in which $a\sim$ 0.29, 0.37, 0.39, 0.38, 0.34 and $b\sim$ 0.24, 0.13, 0.09, 0.07, 0.07 respectively.

Comparing the results in fig.7 with the results in fig.6, we find that the slope of the fitted line in fig.7 is closely related to the occurrence of the giant component. As the size of the largest component $S_{max}$ is quite small, the slope of the fitted line is steep, which indicates that the change of $\xi_0$ can greatly affect $f_C$ as the agents are scattered. As $S_{max}$ has a sharp increase, the slope of the fitted line obviously becomes gentle. As $S_{max}\sim N$, the change of $\rho$ has little effect on the change of $S_{max}$, and accordingly the slope of the fitted line no longer changes with the rise of $\rho$. Such results indicate that, in the present model, the value of b in equation $f_C=a\xi_0^{-b}$ contains the information of evolutionary patterns. We may effectively figure out the size distribution of the components from the slope of the fitted line $f_C(\xi_0)$.

The above simulation results suggest that, as to the decrease of cooperation, the role of increasing $\xi_0$ is similar to the effect of increasing $\rho$. The change of cooperation in the present model should come from the change of the collisions between the agents. Both increasing $\xi_0$ and increasing $\rho$ can effectively increase the collisions between the agents with different strategies, which makes it easy for the defectors to exploit the cooperators and accordingly leads to the decrease of cooperation.

\section{Theoretical analysis}\label{sec:analysis}
\subsection{\label{subsec:levelA}Relationship between extinction threshold $r_L$ and average size $\overline n$ of interaction groups}

In the present model, due to the purposive movement and free-floating of the agents, the interactive partners in the competing group vary with time, which will lead to similar results in well-mixed populations, where all the agents are possible to be chosen as interactive partners in the competing process. In the following, we make a mean field analysis of the replicator dynamics and give an approximation of the extinction threshold $r_L$ in the present model.

According to the payoff function, in a randomly chosen group, a defector always gets a higher payoff than a cooperator. But in mean field analysis, the system does not evolve according to such payoffs but the averaged payoffs for cooperators or defectors which are obtained by averaging over all groups. In the present model, not all the interaction groups have the same size. In theoretical analysis, we take the average size of the interaction groups as the interaction group size, which satisfies $\overline n=9\rho$.

Suppose in the well-mixed population, the frequencies of cooperators, defectors and loners are $f_C$, $f_D$ and $f_L$ respectively, which satisfy the condition $f_C+f_D+f_L=1$. As that in ref.\cite{hauert1}, the average payoffs of defectors, cooperators and loners are

\begin{equation}
P_D=\sigma f_L^{\overline n -1}+r\frac{f_C}{1-f_L}[1-\frac{1-f_L^{\overline n}}{\overline n(1-f_L)}] ,
\end{equation}

\begin{equation}
P_C=P_D-(r-1)f_L^{\overline n -1}+\frac{r}{{\overline n}}\frac{1-f_L^{\overline n}}{1-f_L}-1 ,
\end{equation}

\begin{equation}
P_L=\sigma .
\end{equation}
As the multiplication rate $r$ increases to the extinction threshold $r=r_L$, the loners become extinct and we get $f_L=0$ and $f_C+f_D=1$. In such a case, the payoffs of defectors and cooperators become

\begin{equation}
P_D=\frac{\overline n-1}{\overline n} r_L f_C ,
\end{equation}

\begin{equation}
P_C=\frac{\overline n-1}{\overline n} r_L f_C+\frac{1}{\overline n}r_L-1 .
\end{equation}
On the condition that $P_C=P_D$, where cooperators and defectors can coexist, we obtain the extinction threshold

\begin{equation}
r_L=\overline n.
\end{equation}

According to the above equation, the extinction threshold $r_L$ in the present model is only related to the average size of interaction groups. The rise of population density will lead to the rise of the average group size and thereafter the increase in the extinction threshold. For example, as population density $\rho$ increases from $\rho=0.3$ to  $\rho=0.8$, we can estimate that $r_L$ should increase from 2.7 to 7.2. Compared with the simulation results in fig.1, it is found that, only for an intermediate $\rho$,  the theoretical value of $r_L$ is in accordance with the simulation result. For a small $\rho$,  the theoretical value of $r_L$ is smaller than the simulation result. For a large $\rho$,  the theoretical value of $r_L$ is greater than the simulation result. Such a difference between the mean field analysis and the simulation data may come from the dynamic connectivity between the agents. The above theoretical analysis is only a rough approximation for $r_L$. It has been found that the effect of individual movements can not be handled within the mean-field analysis\cite{szabo5}. That is the reason why the effect of the moving-probability is omitted in our analysis. How to give an accurate approximation for $r_L$ is still an open question for future studies.

\subsection{\label{subsec:levelB}Relationship between free-floating probability $\xi_0$ and levels of cooperation}

The simulation results show that the free-floating of the agents has great impact on the change of cooperation. In the following, by theoretical analysis, we will give a picture of what may be the possible reasons for the occurrence of such an impact.

In the OPGG, the frequencies of cooperators, defectors and loners are determined by the payoffs to different strategies. In the well-mixed case, the payoff of each agent is determined by the group size and the status of the agents in the same group. The larger the group size, the more possible the immediate interactions between the agents. Just like that in the well-mixed case, in a mobile environment, although all the agents are on the sites of a network, they are possible to meet each other within a period of time. The faster the free-floating of the agents, the more possibly the agents meet each other within the period of time. Therefore, from the view point of the probability of meeting between the agents, the effect of increasing the speed of free-floating is the same as the effect of increasing the group size. In the following, we will firstly give a functional relation between the group size and the free-floating probability, and then give a theoretical analysis of the relationship between the free-floating probability $\xi_0$ and the levels of cooperation.

In the present model, because all the agents are arranged on the sites of a square lattice with Moore neighborhood, the number of agents in the same group should be less than or equal to 9. Considering the effects of increasing the group size and the speed of free-floating on the increase of meeting probability, we define the boundary conditions of $\xi_0$ and $n$ as: for $\xi_0\sim 0$, $n\sim 9\rho$; for $\xi_0\sim 1$, $n\sim 9$. Therefore, the following functional relation between $n$ and $\xi_0$ is adopted,

\begin{equation}
n(\rho,\xi_0)=9[\rho+(1-\rho)^\frac{1}{\xi_0}].
\end{equation}

Just like that in a well-mixed case\cite{hauert1,hauert2}, for predefined $\rho$ and $\xi_0$, the difference in the payoff between a defector and a cooperator satisfies the equation

\begin{equation}
P_D-P_C=1+(r-1)f_L^{n(\rho,\xi_0) -1}-\frac{r}{{n(\rho,\xi_0)}}\frac{1-f_L^{n(\rho,\xi_0)}}{1-f_L}.
\end{equation}
In the steady state, the payoffs of all the agents should be the same, $P_C=P_D=P_L=\sigma$. Therefore, the above equation becomes

\begin{equation}
1+(r-1)f_L^{n(\rho,\xi_0) -1}-\frac{r}{{n(\rho,\xi_0)}}\frac{1-f_L^{n(\rho,\xi_0)}}{1-f_L}=0.
\end{equation}
For $0<f_L<1$, the solution of the above equation is the same as the solution of the following equation

\begin{equation}
(1-f_L)+(r-1)(1-f_L)f_L^{n (\rho,\xi_0)-1}-\frac{r}{{n(\rho,\xi_0)}}(1-f_L^{n(\rho,\xi_0)})=0.
\end{equation}
Suppose

\begin{equation}
F(f_L,n(\rho,\xi_0))=(1-f_L)+(r-1)(1-f_L)f_L^{n(\rho,\xi_0) -1}-\frac{r}{{n(\rho,\xi_0)}}(1-f_L^{n(\rho,\xi_0)}).
\end{equation}
For the three cases of $f_L=0$, $f_L=1$ and $f_L\to 1$, we obtain

\begin{equation}
F(0,n(\rho,\xi_0))=1-\frac{r}{n(\rho,\xi_0)},
\end{equation}

\begin{equation}
F(1,n(\rho,\xi_0))=0,
\end{equation}

\begin{equation}
F(f_L,n(\rho,\xi_0))\approx (1-r)(n(\rho,\xi_0)-1)(1-f_L)^2.
\end{equation}

From the equations of (18), (19) and (20) we find that, within the range of $r>2$ and $n(\rho,\xi_0)>r$, as $f_L$ changes from 0 to 1, the value of $F(f_L,n(\rho,\xi_0))$ changes from $F(f_L,n(\rho,\xi_0))>0$ to $F(f_L,n(\rho,\xi_0))<0$. As we check the sign of $\frac{\partial ^2F(f_L,n)}{\partial f_L^2}$ , we find that it changes at most once within the range of $0<f_L<1$. Such results indicate that, within the range of $0<f_L<1$, there exists a single solution $f_L^{tr}$ for equation (16). For $f_L>f_L^{tr}$, the payoff of cooperators is less than the payoff of defectors. For $f_L<f_L^{tr}$, the payoff of cooperators is greater than the payoff of defectors.

The above theoretical analysis indicates that, in the present model, increasing $\xi_0$ will lead to the increase in $f_L$ and accordingly the decrease of $f_C$, which is in accordance with the simulation results found in Fig. 3.

It should be noted that, the above mean field analysis can not accurately predict the cyclic behavior in the dynamic network and the lattice with moving agents. Therefore, the present theoretical analysis is only a rough approximation and the corresponding equations are borrowed from those in a static network\cite{hauert1,hauert2}. The strong relevance of local structures needs a generalized mean-field approximation. A qualitatively correct prediction of the cyclic behavior in the dynamic network and the lattice with moving agents is still an open question for future studies.

\section{Summary}\label{sec:summary}

When facing the choice to win everything or get nothing in a public goods game, it is reasonable for the individuals to drop out of the game and enjoy a small but fixed gain, which yields the rock-scissors-paper cycles of different strategies in the optional public goods game. The oscillatory cooperation in the optional public goods game displays the sustainability of cooperation, but it does not tell us on which conditions the system will evolve to slow oscillations and different levels of stable cooperation can be reached. By incorporating population density and individual mobility into the original OPGG, it is found that both the stability and the improvement of cooperation are connected to the degree of crowdedness and the speed of free-floating of competitive agents.

For low population density and slow free-floating of competitive agents, the departure of dissatisfied agents from competing groups makes it easy for scattered cooperators to agglomerate into cooperator clusters, which results in the expansion of cooperation. The agents who stay on the original sites are more possible to form defector clusters, which will finally be doomed by the random-moving loners. For low population density and fast free-floating of competitive agents, the exchange of neighbors often takes place and it is easy for defectors to invade the cooperative clusters. Numerous small components where cooperators and defectors coexist occur, and the levels of cooperation decrease. For high population density, because of the percolation effect, nearly all the agents are merged into the same component. The defectors uniformly expand into cooperator territory and the evolutionary dynamics in the original OPGG is recovered. The relationship between the levels of cooperation and the free-floating probability is found.

The simulation results in the present model are quite similar to those in ref.\cite{wang5,wang6}, where the effects of different population densities on the evolution of cooperation are studied depending upon the prisoner's dilemma, the snowdrift, the stag-hunt and the public goods game. In a static network with no moving agents, the optimal population density, with which an optimal level of cooperation can be reached, has been found to be related to the percolation threshold.

The present model indicates that the existence of different levels of cooperation in real world should be related to the environmental conditions, including population density and individual mobility. In the future, similar environmental conditions should be considered in the game models with continuous strategy spaces and the generalization of these conditions is a favorite of ours.

\section*{Acknowledgments}
This work is the research fruits of the Humanities and Social Sciences Fund sponsored by Ministry of Education of China (Grant No. 10YJAZH137), Natural Science Foundation of Zhejiang Province (Grant No. Y6110687), Social Science Foundation of Zhejiang Province (Grant No. 10CGGL14YB) and National Natural Science Foundation of China (Grant Nos. 10805025, 11175079, 70871019, 71171036, 71072140).



%% References


%%\bibliographystyle{model1-num-names}
%%\bibliography{<your-bib-database>}



\end{document}